\documentclass[doublecol]{epl2} 
\usepackage{amssymb,amstext,amsmath,hyperref,graphics,epstopdf}

\title{Random matrix approach to multivariate categorical data analysis}
\shorttitle{Random matriix approach for categorical data}

\author{Aashay Patil \and M. S. Santhanam}
\shortauthor{Patil and Santhanam}

\institute{Indian Institute of Science Education and Research, 
Dr. Homi Bhabha Road, Pune, 411008, India.}

\pacs{5.40.-a}{Random processes}

\abstract{
Correlation and similarity measures are widely used in 
all the areas of sciences and social sciences.
Often the variables are not numbers but are instead
qualitative descriptors called categorical data.
We define and study similarity matrix, as a measure of similarity,
for the case of categorical data.
This is of interest due to a deluge of categorical data, such as
movie ratings, top-10 rankings and data from social media, in the
public domain that require analysis.
We show that the statistical properties of the spectra of similarity
matrices, constructed from categorical data, follow those from random matrix theory.
We demonstrate this approach by applying it to
the data of Indian general elections and sea level pressures in North Atlantic ocean.}

\begin{document}
\maketitle

\section{Introduction}
\label{introchap}
Study of correlations is an integral part of almost every branch of science and social science.
Correlated systems and phenomena, such as in non-equilibrium systems, present a rich variety
of behaviour not normally seen in uncorrelated systems. Global climate patterns depend on
the spatial and temporal correlations among atmospheric variables \cite{frb}, correlations in
the stock market records indicate clustering of stocks and indices \cite{b.d,b.z,b.e,b.y}, 
correlations among EEG channels might indicate health of the subject \cite{b.l,eeg1}.
In computer science, correlations are an integral part of most clustering algorithms \cite{cluster}.
In these examples, the object of central interest is the same-time correlation
function $\langle x(t) y(t) \rangle$ for two stationary stochastic processes
$x(t)$ and $y(t)$ with zero mean.
The processes $x_t$ and $y_t$ could be measured data or
generated through simulations. When $N$ variables $x_i(t), i=1,2,...N$ are present, the
correlation matrix $C$ is the appropriate generalisation in which any matrix element
$c_{ij}$ represents the correlation between the variables $x_i(t)$ and $x_j(t)$ \cite{stat}.
It must be noted that singular value decomposition, empirical
orthogonal functions, Karhunen-Loeve decomposition are all variants
of this correlation matrix approach.

Random matrix theory (RMT) \cite{b.b} has emerged as an important tool to understand
the spectra of correlation matrices \cite{eeg1}.
It is by now well established that the spectra of empirical
correlation matrix, for most part, is well described by random matrix results 
\cite{b.d,b.z,b.e,b.y,b.l,eeg1,eeg2,epl,mda}.
Deviations from random matrix behaviour indicate the presence of significant information \cite{b.e}
that cannot be explained purely by assumptions of randomness in matrix elements.
All these methods and analysis, based
on correlations and RMT, depend on the variables $x_i(t)$ being a series of
numbers, representing some possibly stochastic phenomena.

The main objective of this paper is to analyse a measure of association or
similarity for multivariate data sets that are not numbers but discrete qualitative indicators.
Movie ratings and top-ten rankings are some examples of qualitative indicators. Even more 
challenging cases arise when discrete indicators cannot be ranked in any
numerical order. For instance, the
responses in an opinion poll cannot be assigned any meaningful ranking order.
All such data sets are called categorical data \cite{catdata}. In the context of deluge of
data of various kinds available in the
public domain over the internet, it is imperative to look for methods to effectively analyse
categorical data. One important application is in the analysis of data from social media such as
facebook posts, twitter updates, blogs etc., which are mostly not in numerical form.
Social media analysis is now widely used by corporates and even governments to 
understand the public perception of their brand value, products and services.
Hence, computing measures of association with such non-numerical data is often necessary.
Recently, random walks and network theory have been used for computing such measures \cite{plos}.
In contrast, here we develop a statistical technique that is analogous to correlation matrix formalism
and apply RMT tools.

Generally, multivariate empirical data is highly noisy and redundant.
Thus, it is important to separate the information content
from noise components. To do this, we obtain similarity matrix $\mathbf{S}$ as a multivariate generalisation
of similarity measure. We note that similarity matrix $\mathbf{S}$ is widely applied
in clustering algorithms in computer sciences \cite{cluster} and for classifying genetic data \cite{gene}.
By comparing the statistical properties of spectra of $\mathbf{S}$ with that from an
appropriate ensemble of RMT
we can identify the eigenvalues and the eigenmodes that are random.
The spectral components that deviate from RMT results are not random
and generally contain system-specific information yielding
valuable information about the system.
We apply the formalism to two real-world systems,
(i) analysis of Indian general elections results, (ii) mean sea level pressure over North Atlantic region.

\section{Formalism}
In this section, we introduce the formalism for a similarity measure and its 
multivariate generalisation. We consider time series $x_t$ of categorical data. The
elements of the time series are chosen from $p$ possible objects denoted by numbers $1-p$.
Note that the labels $1-p$ do not affect the value of the measure.
For example, $x_t$ could be the time series of parties winning elections in a city.
If there were only two parties (objects) denoted by 1 and 2 that have won election in that city,
then the time series could take the form, $x_t = 1, 2, 1, 1, 2, 1, 2, 2, 1 ...$.
For the case of two time series $x_t$ and $y_t$ of length $T$, we define the similarity measure as
\begin{equation}
c_{\textnormal{xy}}=\mathcal{N} \displaystyle\sum_{t=1}^{T} w_{t} ~ \delta_{x_{t},y_{t}}
\label{cxy}
\end{equation}
where normalisation constant is $\mathcal{N}=\sum_{t=1}^{T} w_{t}$ and
$\delta_{x_{t},y_{t}}$ is the Kronecker delta ($\delta_{a,b}=1$ if $a=b$, $0$ if $a\ne b$).
In this, $w_i$ are the weights assigned to each data point. In most applications, every
data point is given equal weightage and hence $w_t=1$, for all $t=1,2,3...T$. Clearly, $c_{\textnormal{xy}}=1$ only
if $x_t=y_t, \forall ~t$. If $c_{\textnormal{xy}}=0$, this implies $x_t \ne y_t, \forall ~t$.
If $0 < c_{\textnormal{xy}} < 1$, it indicates that $x_t$ and $y_t$ are dissimilar to varying extents.
Note that $c_{\textnormal{xy}}$ is similar to Jaccard index \cite{jaccard1,jaccard2} used
to measure similarity of finite sample sets.

Next, we consider a multivariate scenario with $N$ variables $x_i, i=1,2,....N$, each being
a time series of length $T$. This can be elegantly handled in matrix notation. Let $\mathbf{D}$ represent
a data matrix with of $T$ rows and $N$ columns. Each column is a time series. 
We define a new operator "$\ast$", through its action on two vectors $\mathbf{a} = (a_1~a_2 \hdots a_T)$
and $\mathbf{b} = (b_1~b_2 \hdots b_T)$, defined as
\begin{equation}
\mathbf{a}^T \ast \mathbf{b} = \delta_{a_1,b_1} + \delta_{a_2,b_2} + \hdots \delta_{a_T,b_T}.
\end{equation}
This is similar to applying element-wise AND logical operation between the two vectors.
Using this operator, the multivariate generalisation of similarity measure is
\begin{equation}
\mathbf{S} = \mathbf{D}^T \ast \mathbf{D}
\label{smat}
\end{equation}
In this form, $\mathbf{S}$ has a structure similar to that of Wishart
matrix $C=D^T D$ in multivariate statistics \cite{sgm}. In particular, $\mathbf{S}$ is also a positive
definite matrix with eigenvalues $\lambda \ge 0$. To study the spectra of $\mathbf{S}$, we
numerically solve the eigenvalue equation $\mathbf{S x}_i = \lambda_i \mathbf{x}_i$ 
and obtain its eigenvalues $\lambda_i$ and the eigenvectors $\mathbf{x}_i$.

\section{Similarity matrix}
\begin{figure}
\centerline{\includegraphics[width=2.7in]{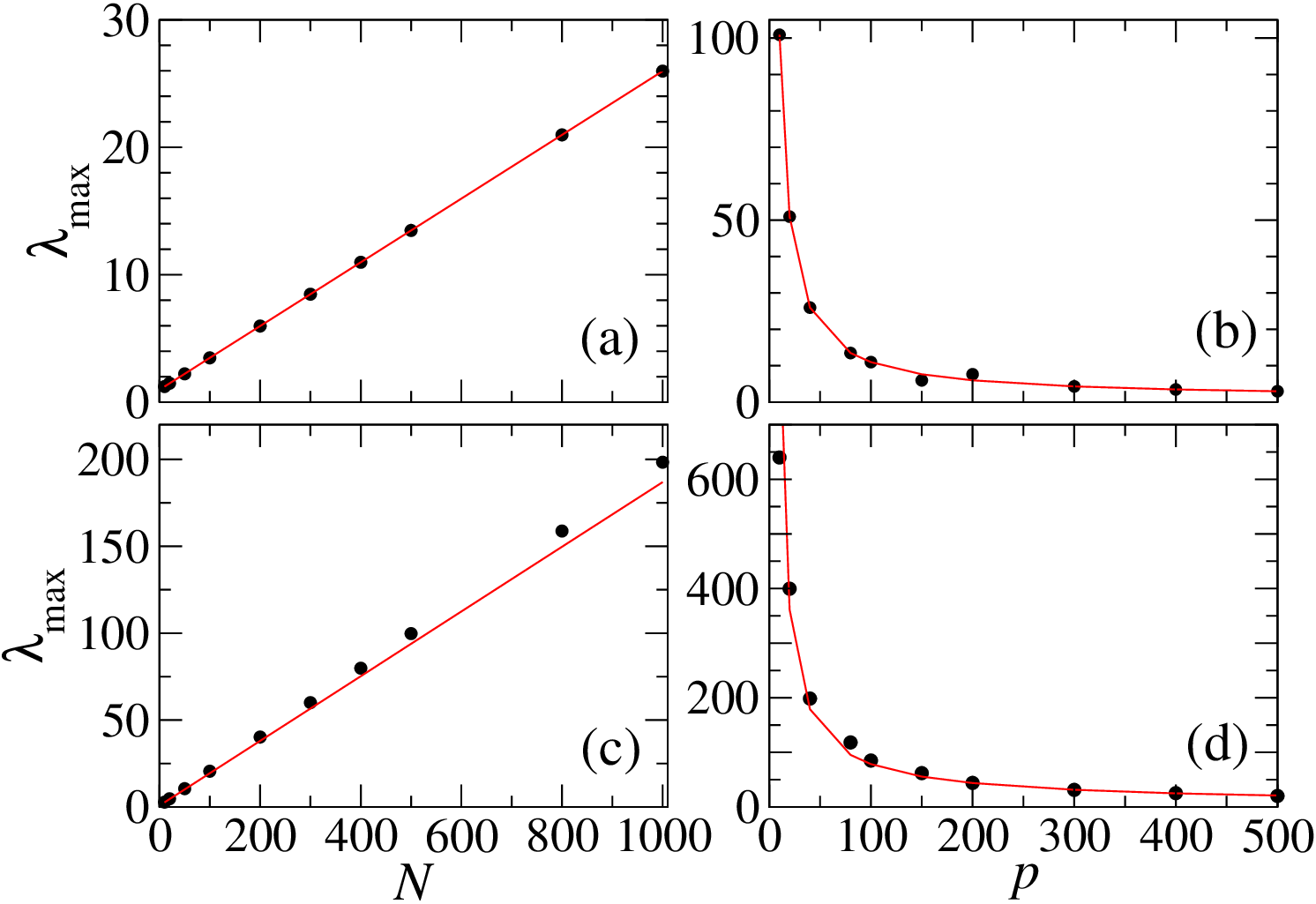}}
\caption{(Colour online) Numerically computed $\lambda_{\textnormal{max}}$ (circles) as a function of number of 
variables $N$ (left)
and number of objects $p$ (right). (a,b) are for uniform distribution
and (c,d) are for geometric distribution of random numbers. The solid lines are the analytical
results in Eqs. \ref{evmaxu} and \ref{evmaxg}.}
\label{lvs}
\end{figure}

This formalism can be illustrated with a simple solvable model. Consider $p$ discrete objects,
labelled 1 to $p$, and $N$ random variables. Each variable $x_i(t), i=1,2,...N$ is a
time series with elements drawn from a {\it discrete} probability distribution $P(\phi)$
for $\phi=1,2,3 \dots p$ and $P(\phi)=0$ otherwise. Then, the elements of $\mathbf{S}$ will be
\begin{eqnarray}
s_{ij} = \frac{1}{T}~\sum_{t=1}^T \sum_{\phi'=1}^{p} \sum_{\phi=1}^{p} P_i(\phi) ~ P_j(\phi') ~
                  \delta_{\phi,\phi'}
\label{smatele}
\end{eqnarray}
Note that $s_{ij}$ turns out to be some function of $p$.
Clearly, by construction, the diagonal elements are $s_{ii}= (1/T) \sum_t \sum_{\phi} P_i(\phi) = 1$.
In the limit $T \to \infty$, $s_{ij}$ would have converged and we get $\mathbf{S}$ to be a matrix of
order $N$ and of the form
\begin{equation}
\mathbf{S}=\begin{pmatrix}
  1 & a & a & ...  \\
  a & 1 & a & ...   \\
  a & a & 1 & ... \\
  : & : & : & 1   \\
\end{pmatrix} .
\label{smatrix}
\end{equation}
The off-diagonal elements are $s_{ij} = a = a(p)$. The eigenvalues of $\mathbf{S}$ can
be analytically obtained. There are only two distinct eigenvalues
\begin{equation}
\lambda_{max} = 1+(N-1)a, \;\;\; \mbox{and} \;\;\; \lambda_1=1-a.
\label{evmax}
\end{equation}
This simple estimate shows that $\lambda_{max}$ is the dominant eigenvalue
and the other eigenvalue is $N-1$ fold degenerate.
We also note that the normalised eigenvector corresponding to $\lambda_{max}$ is
\begin{equation}
(1/\sqrt{N}) (1 ~1 ~1 .....1 ~1).
\label{domevec}
\end{equation}

Now, we can apply this formalism to the case in which the time series $x_i(t)$ are drawn from 
a discrete uniform distribution of the form
\begin{equation}
P(\phi) = \frac{1}{p}, \;\;\;\;\;\; \phi = 1,2 \hdots p.
\label{unidist}
\end{equation}
Then, using Eq. \ref{smatele}, the elements of similarity matrix is $s_{ij}=a=1/p$ for all $i\ne j$.
Then, the eigenvalues are
\begin{equation}
\lambda_{max} = 1+(N-1)/p ~~~~\mbox{and} ~~~~\lambda_1=1-1/p.
\label{evmaxu}
\end{equation}
Next, we consider geometric distribution given by,
\begin{equation}
P(\phi) = (1-q)^\phi q, \;\;\;\;\;\; \phi = 1,2,3 \hdots,
\end{equation}
where $0 < q \le 1$. Note that unlike the uniform distribution (Eq. \ref{unidist}),
the geometric distribution has infinite support. Hence we choose $q$ such that
$\phi=1,2,3 \dots p$ such that $1-\sum_{\phi=1}^p P(\phi)< 10^{-4}$.
Then, $a \approx q/(2-q)$. Using an empirical relation we obtained $q \approx 6/p^{0.8}$,
we get $s_{ij}=a \approx 3/(p^{0.8}-3)$. Then, the eigenvalues are
\begin{equation}
\lambda_{max} \approx 1 + \frac{3 (N-1)}{p^{0.8}-3} ~~~~\mbox{and} ~~~~\lambda_1 \approx 1 - \frac{3}{p^{0.8}-3}.
\label{evmaxg}
\end{equation}
We will use these results as benchmarks to compare with the spectra computed from random similarity matrix.

\section{Random similarity matrix}
In this section, we will study the spectra of random similarity matrix $\mathbf{S_{R}}$ 
in detail. In particular, we compare the statistical properties of $\mathbf{S}$ 
with those obtained from the random similarity matrix. We define random similarity matrix 
$\mathbf{S_R}$ for the case of $p$ objects (labelled by integers 1 to $p$) and $N$ variables
as follows. Let $\mathbf{D_R}$ be a matrix
whose elements are independent and identically distributed integers in the range $[1-p]$
drawn from a discrete probability distribution function.
Then, $\mathbf{S_R} = \mathbf{D_R}^T \ast \mathbf{D_R}$ is the random similarity matrix
of order $N$.

First, we look at how the number of objects $p$ and number of variables $N$ affect
the spectrum of $\mathbf{S_R}$.
We consider $p=40$ and $p=400$ objects with $N=1000$ variables and length of time series
being $T=2000$. All the simulation results (solid circles in Fig 1(a-d)) have been averaged over
100 realisations of appropriate similarity matrix.  Fig. \ref{lvs}(a,b) shows the variation of
$\lambda_{max}$ as a function of number of variables $N$ and number of objects $p$ for the
case of uniform distribution. Surprisingly, the value of $\lambda_{max}$ predicted by
Eq. \ref{evmaxu}, shown as solid line in this figure, holds good even when the elements of
$\mathbf{S_R}$ are noisy due to finite length of time series.
In Fig \ref{lvs}(c,d) shows $\lambda_{max}$ for the case of geometric distribution.
In this case, the number of objects $p$ is approximate and yet the semi-analytical estimate
for the dominant eigenvalue (Eq. \ref{evmaxg}) is in good agreement with the simulated results.
In general, $\lambda_{max}$ decreases with $p$ because
as the number of objects increases, the probability that two time series will have
some common objects decays. For finite number of objects, this decay can be
approximated as $p^{-1}$ for uniform and $p^{-0.8}$ for geometric distribution of random
numbers. In the limit $p\to \infty$, there is only one distinct
eigenvalue $\lambda=1$ and it is $N$-fold degenerate.

\subsection{Eigenvalue Density}
\begin{figure}
\centerline{\includegraphics*[width=2.5in]{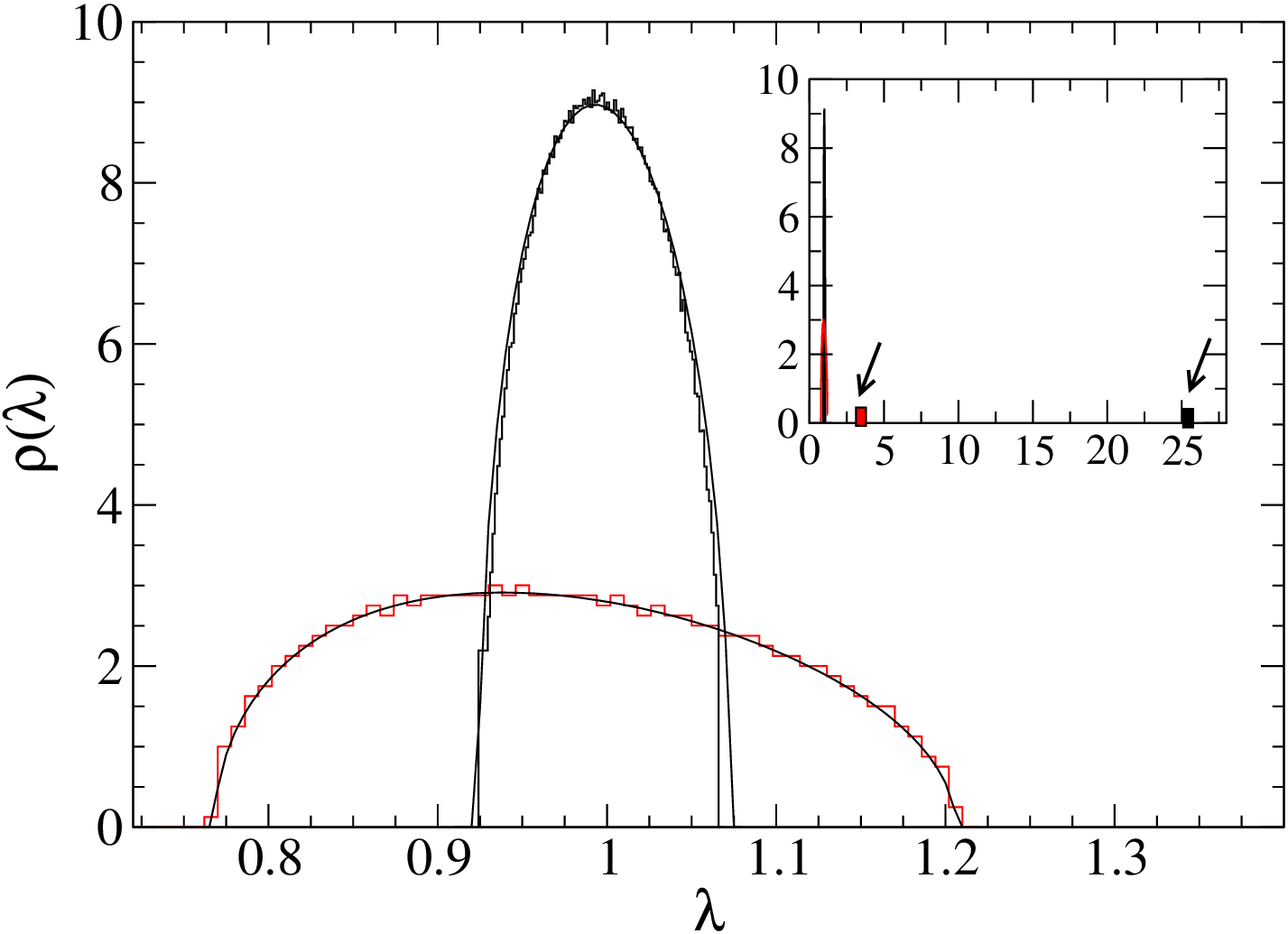}}
\caption{(Colour online) Eigenvalue density for $p=40$ and $p=400$. The histograms are from simulations and 
the solid curves represent Eq. \ref{evdenw}. The inset shows part of the main graph
to focus on the dominant eigenvalues (indicated by arrows) which are far
from the bulk of eigenvalues.}
\label{evden}
\end{figure}

\begin{figure}
\centerline{\includegraphics*[width=2.5in]{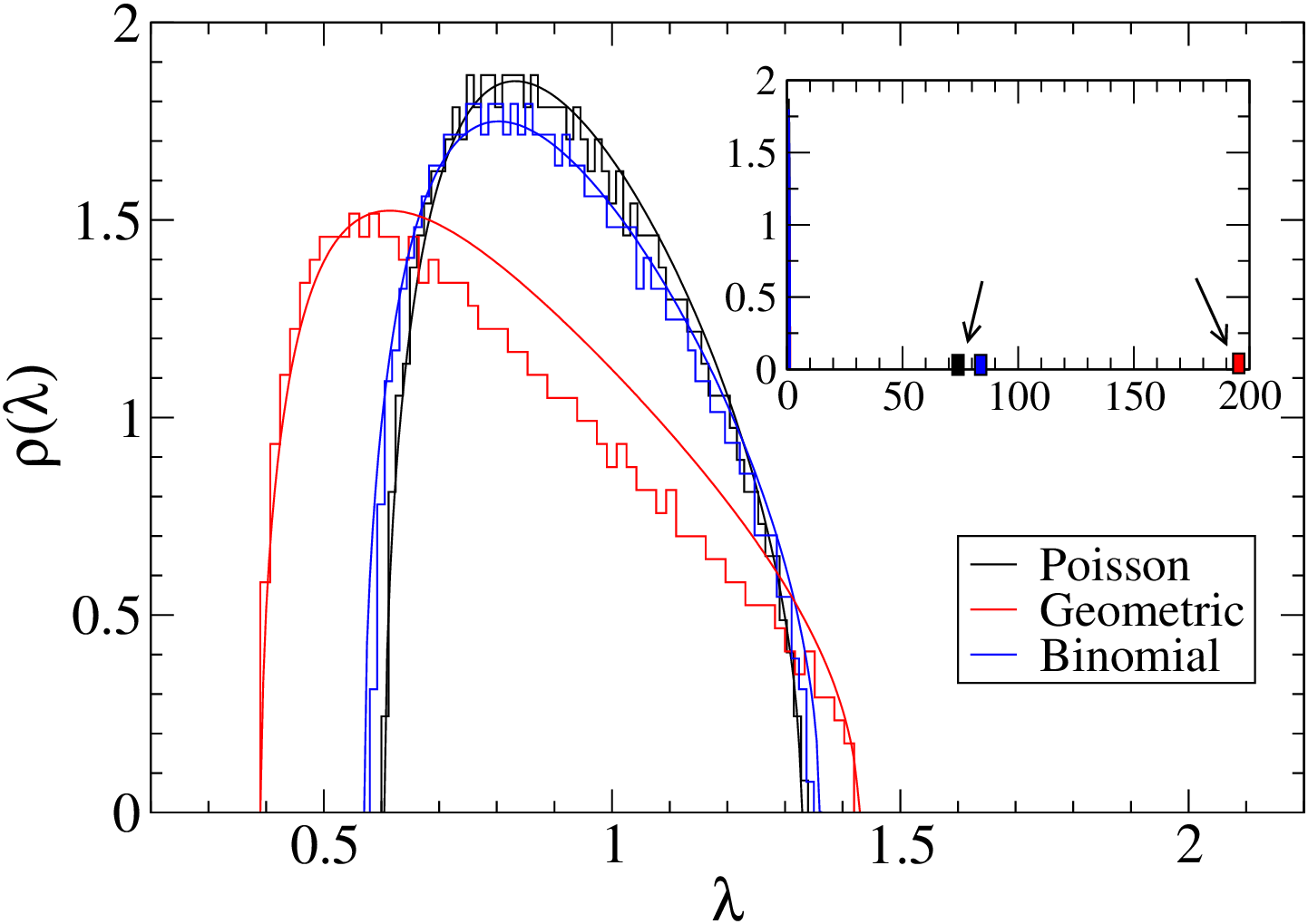}}
\caption{(Colour online) Eigenvalue density for $\mathbf{S_R}$ with random variables from
discrete distributions, Poisson, Binomial and Geometric. Histograms are from simulations and
solid lines are fitted using Eq. \ref{evdenw}. Inset shows $\rho(\lambda)$
focussing on the dominant eigenvalues (highlighted by arrows).}
\label{evden_bin}
\end{figure}

We study two quantities that characterise the eigenvalues of $\mathbf{S_R}$, namely, eigenvalue density
and spacing distribution.
The mean density of eigenvalues is defined by
\begin{equation} 
\rho(\lambda)=\sum_i \delta(\lambda-\lambda_i),
\end{equation}
where $\delta(.)$ is the Dirac-delta function.
Given that Eq. \ref{smat} has a structure
similar to that of Wishart matrix, it is reasonable to expect the density of
eigenvalues for random Wishart matrix $\mathbf{C}=\mathbf{D}^{\textnormal{T}} \mathbf{D}$
to hold good for random similarity matrix as well.
In Wishart case, if $\mathbf{D_R}$ is a $T\times N$ random matrix with uncorrelated column vectors
drawn from a Gaussian distribution with mean $\mu$ and variance $\sigma^2$, then $\rho(\lambda)$,
in the limit $N\to\infty$ and $T\to\infty$
and $Q=\frac{T}{N}\geq1$ is the Marchenko-Pastur law \cite{sgm}
\begin{equation} 
\rho(\lambda)=\frac{Q}{2\pi\sigma^{2}}\frac{\sqrt{(\lambda_{\textnormal{max}}-\lambda)(\lambda-\lambda_{\textnormal{min}})}}{\lambda},
\label{evdenw}
\end{equation}
for $\lambda\in [\lambda_{\textnormal{min}},\lambda_{\textnormal{max}}]$ and $\rho(\lambda)=0$
otherwise. In this, the largest and smallest eigenvalues are
\begin{equation} 
\lambda_{\textnormal{max}/\textnormal{min}}=\lambda_{+/-}=\sigma^{2}(1+1/Q\pm 2\sqrt{1/Q}).
\end{equation}
In the limit $Q=1$, the eigenvalue density leads to the well-known Wigner semi-circle law \cite{wigner}.
We compare Eq. \ref{evdenw} with eigenvalue density computed for random similarity matrix.

The eigenvalue density, for the bulk of eigenvalues, of random similarity matrix 
$\mathbf{S_R}$ is shown in Fig. \ref{evden} and it is well described
by Eq. \ref{evdenw}.
On the other hand, the largest eigenvalue $\lambda_{max}$, highlighted in the
inset of Fig. \ref{evden}, is an order of magnitude larger than all the other eigenvalues.
It stands out from the bulk. This is a unique spectral signature of
random similarity matrix $\mathbf{S_R}$. A matrix such as $\mathbf{S_R}$ that encodes
random correlations, in the spirit of random matrix theory, is not expected to accord special
treatment for any part of the spectrum. Yet, the dominant eigenvalue $\lambda_{max}$ has
a special place in the spectrum.
The $\rho(\lambda)$ for Poisson, Binomial and Geometric
distribution of random numbers shown in Fig. \ref{evden_bin} also display a similar feature for
$\lambda_{max}$. In this case too (Fig. \ref{evden_bin}) the bulk of eigenvalues are reasonably 
consistent with Eq. \ref{evdenw}.
The mild deviation for geometric distribution case in this figure can be attributed to the
approximate nature of the calculation due to its infinite support.

\subsection{Spacing distribution}
\begin{figure}
\includegraphics*[width=3.2in]{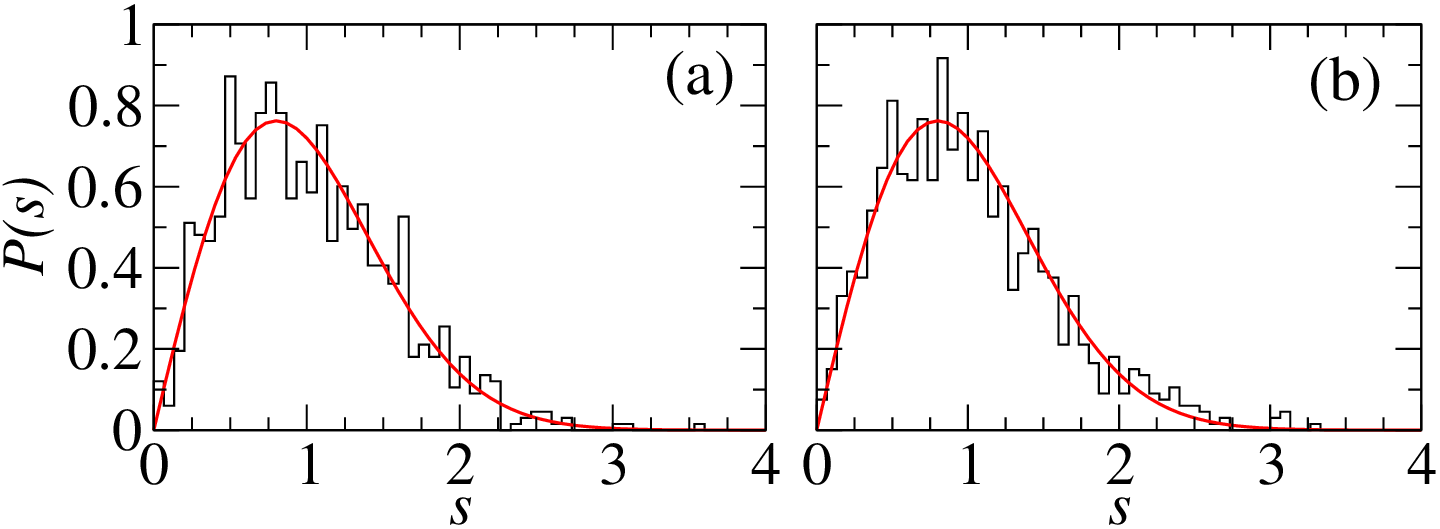}
\caption{(Colour online) Empirical spacing distribution for the eigenvalues obtained from
$\mathbf{S}$ with elements of $\mathbf{D}$ drawn from
(a) uniform distribution and (b) geometric distribution. In both cases,
$p=40, N=1000$ and $T=2000$. The solid curve is Wigner distribution (Eq. \ref{wigdist}).}
\label{sdist}
\end{figure}
In this section, we present results for the spacing distribution of the eigenvalues.
We remove the dominant eigenvalue $\lambda_{max}$ and compute spacing distribution using
all the other eigenvalues in the bulk (see Figs. \ref{evden}-\ref{evden_bin}).
If the eigenvalues of $\mathbf{S_R}$ are represented by $\lambda_i, i=1, 2,\dots N$,
we transform the eigenvalues to obtain 'unfolded' eigenvalues $\epsilon_i, i=1, 2,\dots N$,
The nearest neighbour spacings are defined as $s_i=\epsilon_{i+1}-\epsilon_i$ such
that $\langle s \rangle =1$. Given that $\mathbf{S_R}$ is real symmetric matrix with
random entries, we expect the empirical spacing distribution obtained from the spectra
of $\mathbf{S_R}$ to be
best described by Gaussian Orthogonal Ensemble (GOE) result, the Wigner distribution,
of random matrix theory \cite{b.b}.
Hence, the appropriate result is,
\begin{equation}
P_W(s) = \frac{\pi}{2}s ~e^{-\frac{\pi}{4}s^2}.
\label{wigdist}
\end{equation} 
In Fig. \ref{sdist} we show the computed spacing distribution for the eigenvalues
of $\mathbf{S_R}$ with the matrix elements of $\mathbf{D_R}$ drawn from discrete uniform
and geometric distributions. For both these cases, the spacing distributions follow
the random matrix theory results in Eq. \ref{wigdist}. It must be recalled that
similar results hold good for the spacing distribution of empirical correlation matrices \cite{b.e,msspp}.
We further note that as $T\to \infty$, the matrix elements of $\mathbf{S_R}$ converge
to their true values and the spacing distribution deviates strongly from Eq. \ref{wigdist}.

\subsection{Eigenvector statistics and Information Entropy}

\begin{figure}
\includegraphics*[width=2.7in]{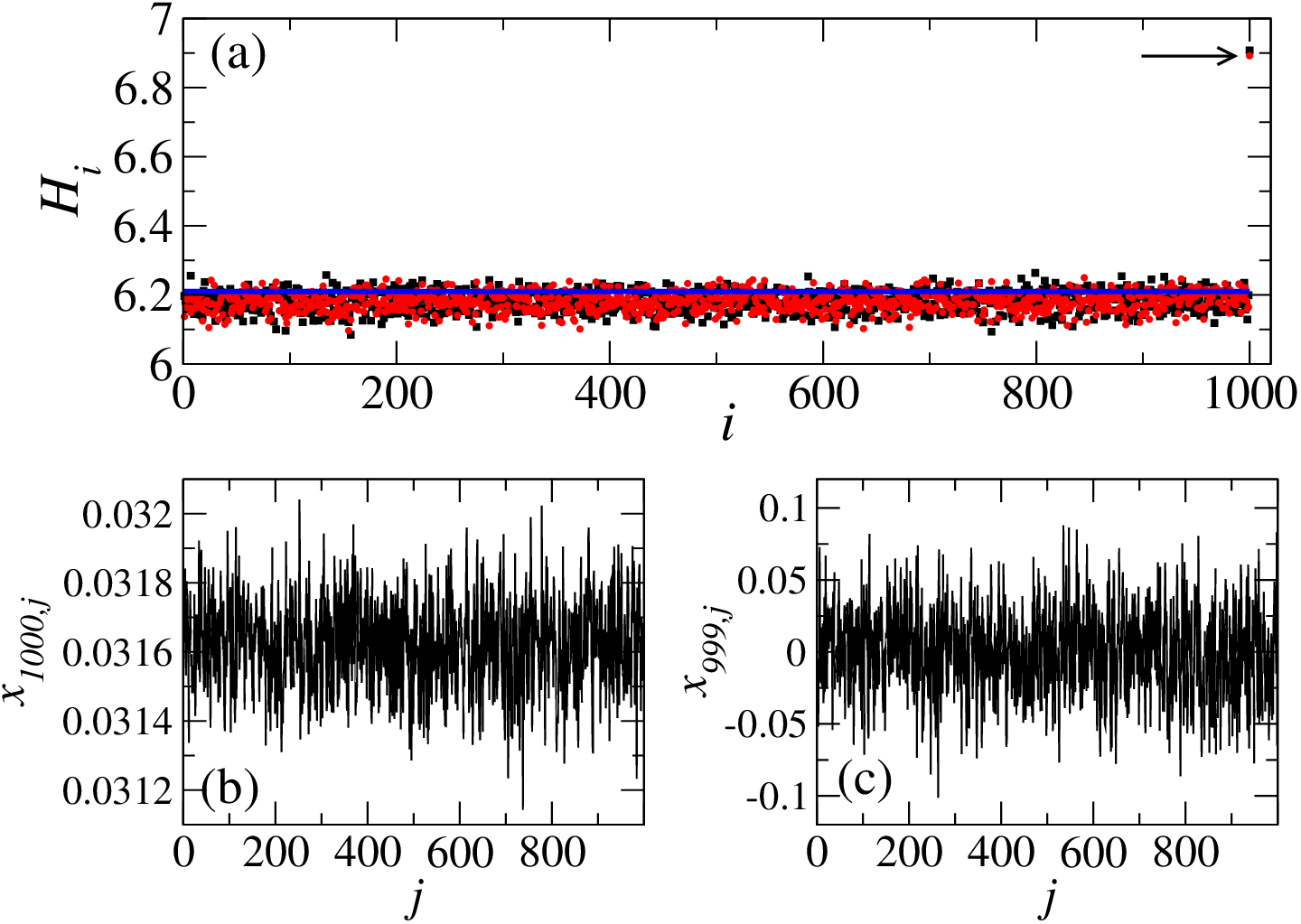}
\caption{(Colour online) (a) Information entropy for the eigenvectors of $\mathbf{S_R}$ for the case of
uniform distribution (black circle) and geometric distribution (red square).
$H_i$ for the dominant eigenvector stands out from the bulk and is highlighted
by an arrow. For the case of $\mathbf{S_R}$ obtained for the uniform distribution case
(b) shows eigenvector of the dominant eigenvalue and (c) shows eigenvector for an
eigenvalue in the bulk.}
\label{infent}
\end{figure}

In this section, we study the properties of eigenvectors $\mathbf{x}_i, i=1,2,...N$ of $\mathbf{S_R}$.
The eigenvectors corresponding to the eigenvalues in the bulk are
Gaussian distributed (not shown here), in accordance with the random matrix results \cite{b.b}. A comprehensive
comparison with random matrix results can be done by computing the information
entropy for the $i$-th eigenvector defined by \cite{b.j}
\begin{equation} 
H_i= -\sum_{j} |x_{ij}|^2 \ln |x_{ij}|^2.
\end{equation}
The corresponding random matrix average for the information entropy is
given by $H^{RMT} \sim \ln (N/2)$ \cite{b.j}, where $N$ is the dimension of the random matrix.
We show the information entropy $H_i$ as a function of eigenvalue index $i$ in Fig. \ref{infent}.
The information entropy $H_i$ for the bulk of eigenvectors
follow random matrix result $H_i \approx \ln 500 = 6.214$, indicated as a blue line. 
As an instance of such an eigenvector in the
bulk, we show in Fig \ref{infent}(c) the 999th eigenvector components $x_{999,j}$. The random
nature of this eigenvector is clearly visible in its oscillations about zero. This behaviour must
be contrasted with the dominant eigenvector (corresponding to $\lambda_{max}$) $x_{1000,j}$ shown
in Fig \ref{infent}(b).
In this case, though the oscillations exist, they are not about zero, i.e, all the components
of this eigenvector have identical phase. This behaviour can be understood based on the fact
that for the simple model in Eq. \ref{smatrix}, obtained in the $T\to \infty$ limit, the dominant
eigenvector has the form shown in Eq. \ref{domevec}. Note that phases of all the components
are identical in Eq. \ref{domevec} as well. For the dominant eigenvector of $\mathbf{S_R}$ the
amplitudes become random but not the phases.
This non-random phases leads to significant deviation from
random matrix average $H^{RMT}$ for information entropy as indicated by the arrow in Fig. \ref{infent}(a).
Thus, deviations from random matrix results imply presence of correlations either in the amplitude
or the phase of the eigenvectors. This, in turn, could be traced to the correlations in
the similarity measure for many variables.

\section{Application}
We apply the formalism to two different data sets,
(i) the data of Indian general elections and
(ii) atmospheric pressure in the region of North Atlantic ocean.
We describe the motivations for choice of these data sets and 
their details below.

\subsection{Elections data}
\begin{figure}
\centerline{\includegraphics*[width=2.8in]{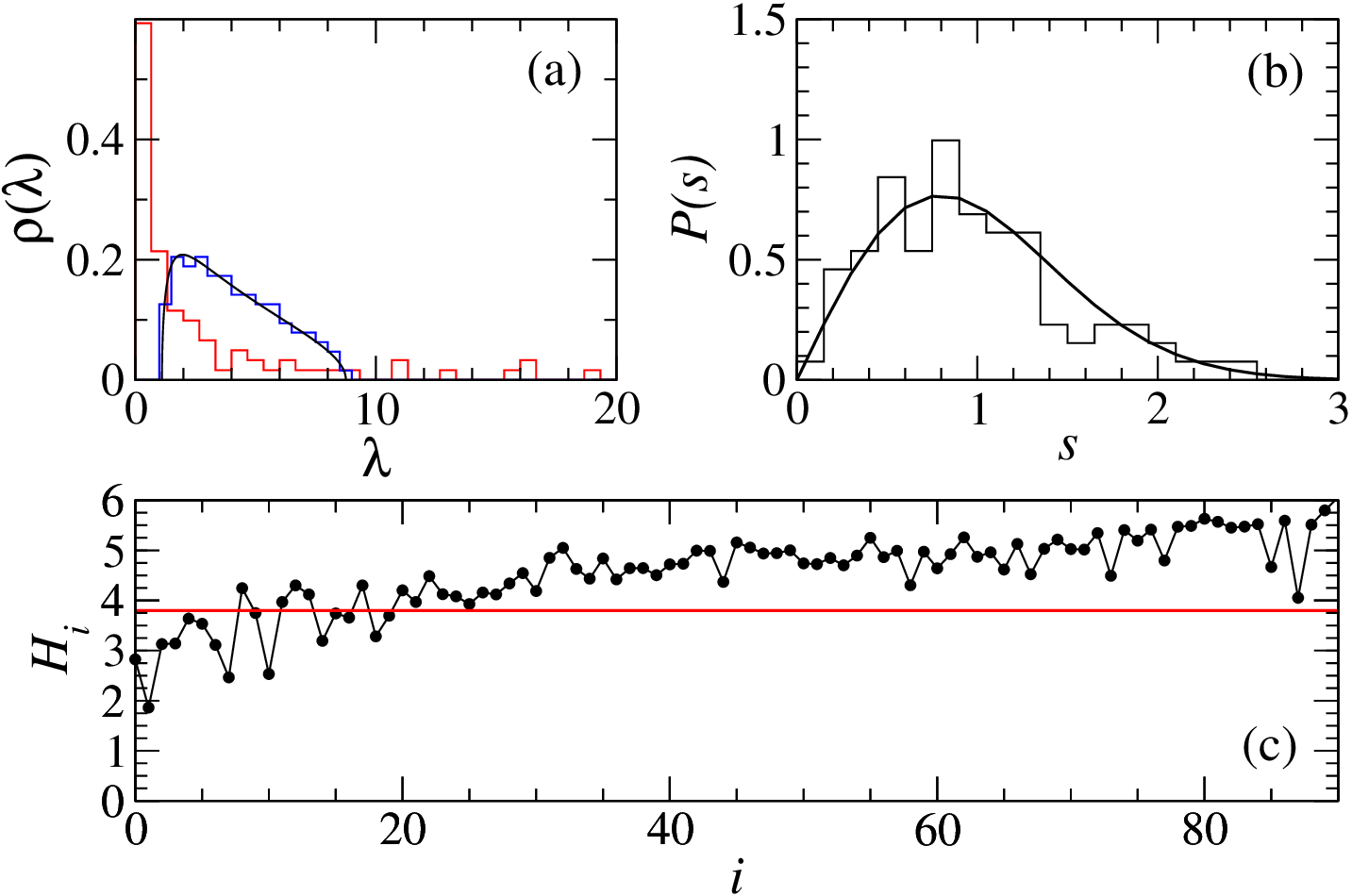}}
\caption{(Colour online) Results from the spectra of $\mathbf{S}$ obtained from data on Indian general
elections. (a) Eigenvalue density for elections data compared against random similarity
matrix, (b) spacing distribution obtained from elections data and (c) information
entropy. See text for details.}
\label{elect}
\end{figure}
The general elections
held in India to elect the lower house of Indian parliament is the largest democratic
exercise of its kind in the world with about 814.5 million people eligible to exercise their right to vote.
These elections elect 543 representatives from as many constituencies to the lower house,
Lok Sabha. For the purposes of our analysis, we identify 19 major political parties that have had
significant representation in the elections held in India since 1984.
These parties form our objects, i.e, $p=19$. For each constituency, the data we employ is a
time series of winning party at seven general elections held during 1984-2004 and hence $T=7$.
The number of variables is the number of constituencies, $N=543$.
The general elections data, dating back to the first one in 1952, is provided by the
Election Commission of India \cite{pol} and all the analysis reported here is based on this data.
In this scenario, the similarity measure
$s_{ij}$ is an index of how close are $i$-th and $j$-th constituencies in terms of the
parties they have elected in the series of general elections. For instance, $s_{ij}=1$
implies that $i$-th and $j$-th constituencies have exactly chosen the same set of
parties in all the general elections.

We note that the length of the time series is small and hence the computed
matrix $\mathbf{S}$ is singular. This is also evident from the fact that out of
543 eigenvalues, only 91 of them are non-zero which form the basis for the
results of eigenvalue statistics presented in Fig. \ref{elect}(a,b).
In Fig. \ref{elect}(a), we show the computed eigenvalue density $\rho(\lambda)$.
We note that unlike in the cases shown in Figs. \ref{evden} - \ref{evden_bin}, many
eigenvalues, both at the lower and upper end, deviate from
random matrix formula (Eq. \ref{evdenw}). Even though the spacing distribution, shown
in  Fig. \ref{elect}(b), largely
follows Wigner distribution there are visible deviations as well. This could be attributed
to poor statistics and to the fact that election data is strongly correlated as well.
This is further corroborated by the deviations in $H_i$ from random matrix results (shown
as red line in Fig. \ref{elect}(c)).

\subsection{Atmospheric pressure data}
\begin{figure}
\centerline{\includegraphics*[width=2.8in]{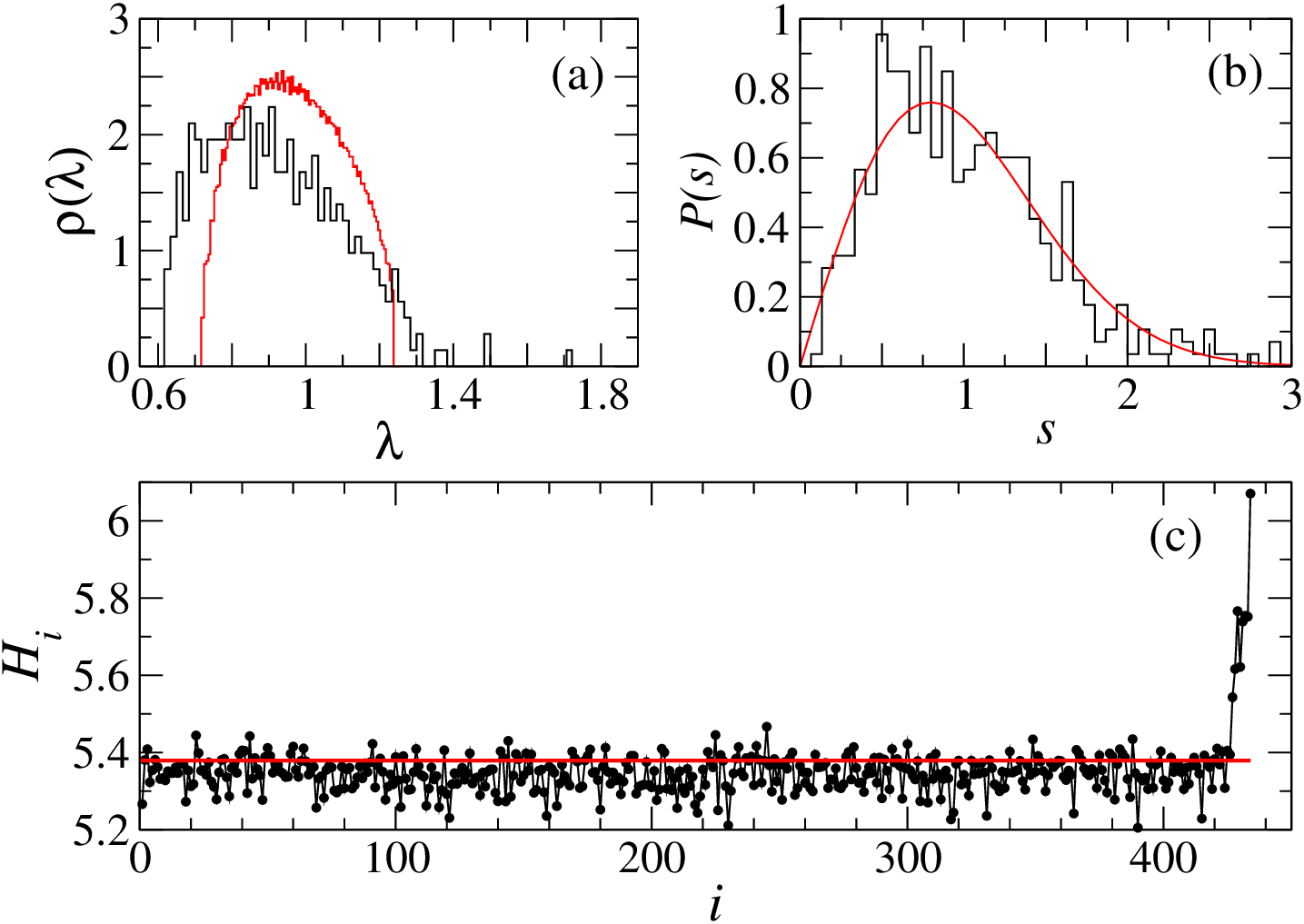}}
\caption{(Colour online) Statistics of spectra of $\mathbf{S}$ obtained from data of sea level
pressure. (a) Eigenvalue density for atmospheric pressure data compared against that from random similarity
matrix, (b) spacing distribution (histogram) obtained from data and Wigner distribution (solid line),
(c) information entropy from data (solid circles) and its RMT average (red line). See text for details.}
\label{atmpress}
\end{figure}
Now, we consider the sea level atmospheric pressure (SLP) data over the North Atlantic
region. This region and in particular this set of data has been well studied in order
to understand a pressure see-saw phenomena called the North Atlantic oscillations.
In contrast with the elections data in which the political parties (objects) are
discrete entities, in this case the SLP values (objects) are continuous.
Notice that the formalism requires the objects to be discrete.
Hence, we discretise the data as follows. If $a_{max}$ and $a_{min}$ represent the
maximum and minimum observed value in the data, we create data intervals of width
\begin{equation}
\Delta=\frac{a_{max}-a_{min}}{p}.
\end{equation}
Suppose a data value $a$ lies in, say, 3rd interval $(a_{min}+3\Delta,a_{min}+4\Delta)$, then the discretised data
corresponding to $a$ is $3$. In signal processing, this technique of mapping continuous
data to a countable set of integers is called quantization \cite{quant}.
By this process, the entire data set is converted into time series
of integers (representing data intervals).
Since the observed data in any measurement is known to be contaminated by
errors and instrumental noise, it is only fitting that intervals of observed values are analysed
instead of the actual values.

We use the NCEP reanalysis data of monthly mean sea level
pressure at 434 grid points on the sea surface for the period 1948-2001 \cite{url}. A correlation matrix
analysis of this data from the point
of view of random matrix theory was reported in Ref. \cite{msspp}.
The data has $N=434$ variables and each variable has time series length of $T=624$.
The number of objects (data intervals) is $p=40$. The similarity index $s_{ij}$ in this
case measures if the variations of sea level pressure at $i$-th and $j$-th geographical locations
are similar. If $s_{ij}=1$, then the discretised data values at these two locations
are identical.

Using this discretised data, we compute matrix $\mathbf{S}$ and its spectra. For comparison
purposes, we also compute the spectra of its random matrix equivalent $\mathbf{S_R}$.
Similar to the case of elections data, the eigenvalue
density shown in Fig \ref{atmpress}(a) (as histogram) displays deviations from that of its random
matrix (shown as red curve) at the lower and upper end. These deviating eigenvalues indicate correlations
or system specific information that cannot be modelled by randomness assumptions.
In this case too, the spacing distribution shown in Fig. \ref{atmpress}(b) agrees with
the Wigner distribution $P_W(s)$. The eigenvectors, corresponding to the  deviating eigenvalues 
in Fig \ref{atmpress}(a),
also display pronounced deviation from random matrix averages. This is seen in Fig. \ref{atmpress}(c)
which shows the information entropy as a function of index of eigenvalue.
The dominant eigenvectors at the top end of the spectrum are known to capture the pressure patterns
that are relevant in atmospheric sciences \cite{msspp}. We also point out that the components of dominant
eigenvector of $\mathbf{S}$, for both elections data and SLP data, have identical phases (not shown here),
in agreement with the result shown in Fig. \ref{infent}(b).

\section{Summary}
We have studied the problem of computing a measure of similarity
for multivariate time series of categorical data, i.e., time series data sets that are not in numerical form.
Such data sets are encountered in many situations in social media, say, as response to
major events or speeches, in the context of stars or recommendations given to
movies or books or other such resources. We construct a similarity matrix $\mathbf{S}$ 
by assembling together the similarity measure $s_{ij}$ between $i$-th and $j$-th
variables. Further, we study the spectra of $\mathbf{S}$ for the case of uncorrelated
categorical data and compare with appropriate random matrix results.
For most part, the spectra of $\mathbf{S}$ follow random matrix theory prescriptions though
the dominant eigenvector deviates due to phase coherence.
The eigenvalues and eigenvectors of $\mathbf{S}$ that deviate from random matrix results
are seen to signify the presence of correlations that cannot be explained by
randomness assumptions that underlie random matrix theory. As an application of this approach,
we use the data on the Indian general elections and atmospheric pressure in the North Atlantic ocean region to study the similarity properties.
The former is an example which lends itself readily to this analysis but in the latter example
the original data is quantized before analysis. Thus, analysis described in this paper 
can be performed on most of empirically available data sets.



\acknowledgments
NCEP Reanalysis data provided by the NOAA/OAR/ESRL PSD, Boulder, Colorado, USA, 
from their web site www.esrl.noaa.gov/psd. The data of Indian general
elections is available from eci.nic.in.

\end{document}